\documentclass[5p,twocolumn]{elsarticle}
\journal{Nuclear Instruments and Methods in Physics Research Section A}
\usepackage[USenglish]{babel}
\usepackage[T1]{fontenc}
\usepackage[utf8]{inputenc}
\usepackage{amssymb} 
\usepackage{amsmath}
\usepackage{amsthm}
\usepackage{amsfonts}
\usepackage{siunitx}
\usepackage{booktabs}

\newcommand{\degree}{\ensuremath{^\circ}}

\begin{document}


\title{Full PIC simulation of a first ACHIP experiment @ SINBAD}

\author[desy,uhh]{W.~Kuropka}
\ead{willi.kuropka@desy.de}

\author[desy,uhh]{F.~Mayet}
\author[desy]{R.~Aßmann}
\author[desy]{U.~Dorda}

\address[desy]{Deutsches Elektronen-Synchrotron DESY, Notkestraße 85, 22607 Hamburg, Germany}
\address[uhh]{Universität Hamburg, Mittelweg 177, 20148 Hamburg, Germany}

\begin{abstract}
In laser illuminated dielectric accelerators (DLA) high acceleration gradients can be achieved due to high damage thresholds of the materials at optical frequencies. This is a necessity for developing more compact particle accelerator technologies. The Accelerator on a CHip International Program (ACHIP) funded by the Gordon and Betty Moore Foundation is researching such devices. DESY Hamburg is part of the collaboration. The dedicated accelerator research facility SINBAD is particularly well suited for DLA experiments at relativistic electron energies. High quality beams and short bunch lengths are anticipated from the ARES linac which is currently under construction at SINBAD.
The aim of the experiment is the injection of a short electron bunch from the ARES linac into a DLA. In this study the results of one of the first possible experiments at the facility are estimated via a combination of particle-in-cell (PIC) and tracking simulations. ASTRA is used to simulate an electron bunch from the ARES linac at a suitable working point. The dielectric part of the setup will be simulated using the PIC code from CST Particle Studio incorporating the retrieved bunch from the ASTRA simulation. The energy spectra of the electron bunches are calculated as would be measured from a spectrometer dipole with and without the laser fields.
\end{abstract}


\maketitle



\section{Introduction}

The Accelerator on a CHip International Program (ACHIP) is a research project funded by the Gordon and Betty Moore Foundation. It aims at the construction of a compact fully laser driven dielectric electron accelerator for radiation generation and atto-second science. Several Universities in Europe and the USA and the national Laboratories PSI, DESY and SLAC are involved \cite{ACHIP}.

This research field gains more attention in recent years as the search for compact particle acceleration technologies continues. Fully dielectric laser driven acceleration structures are foreseen to sustain high acceleration gradients in the \SI{}{\giga\volt\per\m} regime. This is mainly due to the high laser damage thresholds of dielectrics at optical frequencies. The advancement in micro- and nano-fabrication from the semiconductor industry can be leveraged in the manufacturing of these structures limiting production costs. Also the ongoing development in laser technologies promises higher repetition rates and average powers suitable for DLAs \cite{England2014}.

DESY Hamburg contributes to the ACHIP project by granting access to its SINBAD facility and support from the ARD (accelerator research and development) and laser groups. SINBAD is the upcoming dedicated lon term accelerator research and development facility at DESY. At the moment two projects are foreseen to be implemented. The first is ARES (Accelerator Research Experiment at Sinbad), which is based on a linac consisting of an S-band, normal conducting radio frequency gun and traveling wave accelerator structures. The second project is called AXSIS (Frontiers in Attosecond X-ray Science:
Imaging and Spectroscopy \cite{AXSIS}) and aims at X-ray radiation generation via inverse compton-scattering with electrons from THz-driven dielectric lined waveguides \cite{Ulrich1}.

The location of the ACHIP related experiments will be at the ARES linac. It is foreseen to inject already relativistic electron bunches at around \SI{100}{MeV} from the linac into the DLA structures. Its high beam quality and short femtosecond electron bunches are well suited for DLA experiments with small apertures and optical drive laser wavelengths. By driving the DLA with at least \SI{2}{\micro\m} wavelength a single bunch net-acceleration experiment is feasible.

In this work we present a full PIC (CST Particle Studio \cite{CST}) simulation of the DLA interaction with an ASTRA \cite{ASTRA} simulated electron bunch from the ARES linac. The energy spectra with and without the laser field are calculated as anticipated from a dipole spectrometer. The results are compared with a simplified simulation code for DLAs developed in-house in our group \cite{Kuropka2017}.


\section{ARES working point and experimental setup}

\begin{figure}[!htb]
   \centering
   \includegraphics*[width=78mm]{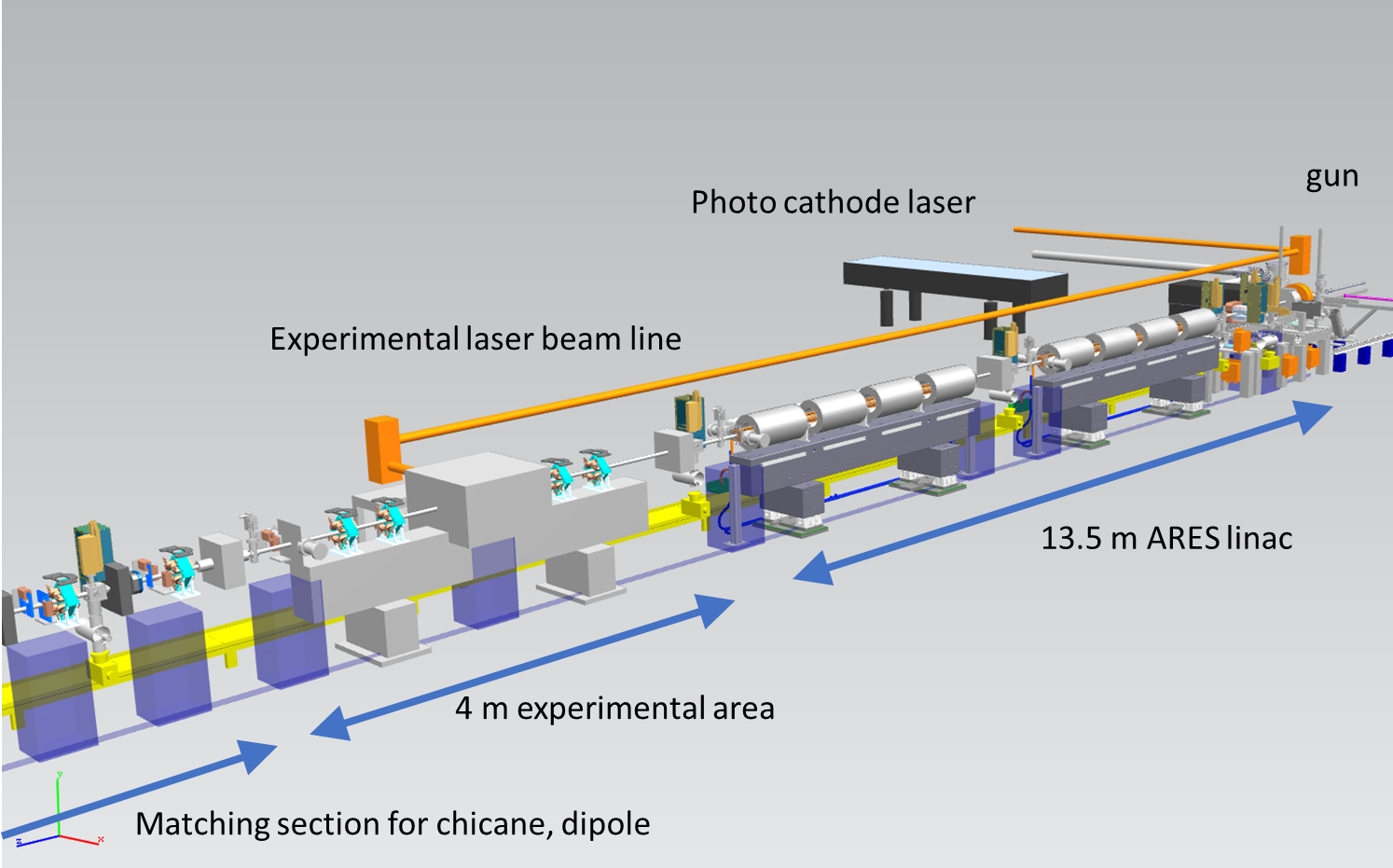}
   \caption{Overview of the ARES linac with reserved space for the experimental area.}
   \label{fig:ARES_Overview}
\end{figure}

The ARES linac consists of an S-band gun and two traveling wave structures (TW). The experimental area will be situated after the second TW shown in figure \ref{fig:ARES_Overview}. For this working point the first TW is used off-crest for velocity bunching yielding a longitudinal focus into the second TW structure where the compressed bunch length is fixed due to reaching high Lorentz-factor. The TWs are equipped with solenoid magnets to focus the beam transversely. With this setup the working point shown in table \ref{tab:ARES_WP} is achievable in simulation.

\begin{table}[hbt]
   \centering
   \caption{Simulated working point at structure center for minimal bunch length using velocity bunching from \cite{Mayet:IPAC2017-WEPVA007}}
	 \vspace{0.1cm}
   \begin{tabular}{ll}
      \toprule
      \textbf{Parameter @ IP} & \textbf{Value}\\
      \midrule
      Charge [pC] & 0.5\\
      Bunch Length [fs, FWHM] & 2.1 (0.3 * $\lambda_{0}$)\\
      E [MeV] & 99.1\\
      $\Delta$E/E [\%] & 0.12\\
      $\sigma _{xy}$ [$\mu$m] & 7.8\\
      $\epsilon _{n,xy}$ [nm] & 105\\
      \bottomrule
   \end{tabular}
   \label{tab:ARES_WP}
\end{table}

For the first experiments the drive laser power for the DLA is split from the ARES cathode laser. A small part of the power is converted to UV via a fourth-harmonic generator. An optical parametric amplifier is used to convert the remaining laser power from \SI{1}{\micro\m} to \SI{2}{\micro\m}. This is feasible due to low charge requirements for this kind of experiment. The cathode laser delivers \SI{1}{\milli\J}. For this setup the split-off and the conversion efficiency of an OPA of around \SI{10}{\percent} only allow for low two digit \SI{}{\micro\J} pulse energy at the \SI{2}{\micro\m} wavelength. Using this wavelength with a period of \SI{6.67}{\femto\second} a net-gain in particle energy is possible since the bunch only covers a limited phase range of around one third of the accelerating field. Other methods can be implemented to achieve net-gain of a bunch train and to reduce arrival time jitter \cite{FrankEAAC1}.

The cathode laser pulse has to be short to achieve the short bunch lengths necessary, about \SI{120}{\femto\second}. The laser pulse duration at the DLA is identical to the duration at the cathode. The temporal and spatial shape of the laser pulse are Gaussian. Table \ref{tab:Laser} shows the collected laser parameters. The maximum electric field amplitude in the channel for the given pulse energy is around \SI{4}{\giga\V\per\m}. These parameters are still well below the anticipated damage threshold of fused silica \cite{Jia2006}. To have the maximum possible energy gain the interaction length is elongated by introducing a \SI{45}{\degree} pulse front tilt to the DLA drive pulse. The resulting interaction length is \SI{100}{\micro\meter}. The DLA for the simulation is assumed with this length containing ca. 50 periods.


\begin{table}[hbt]
   \centering
   \caption{laser pulse parameters}
	 \vspace{0.1cm}
   \begin{tabular}{ll}
      \toprule
      \textbf{Parameter @ IP} & \textbf{Value}\\
      \midrule
      Wavelength [$\mu$m] & 2\\
      $\sigma _{t}$ [fs] & 120\\
      Energy [$\mu$J] & 14\\
      $\sigma _{x}$ [$\mu$m] & 50\\
      $\sigma _{y}$ [$\mu$m] & 10\\
      \bottomrule
   \end{tabular}
   \label{tab:Laser}
\end{table}


\section{PIC simulation}

\begin{figure}[!htb]
   \centering
   \includegraphics*[width=78mm]{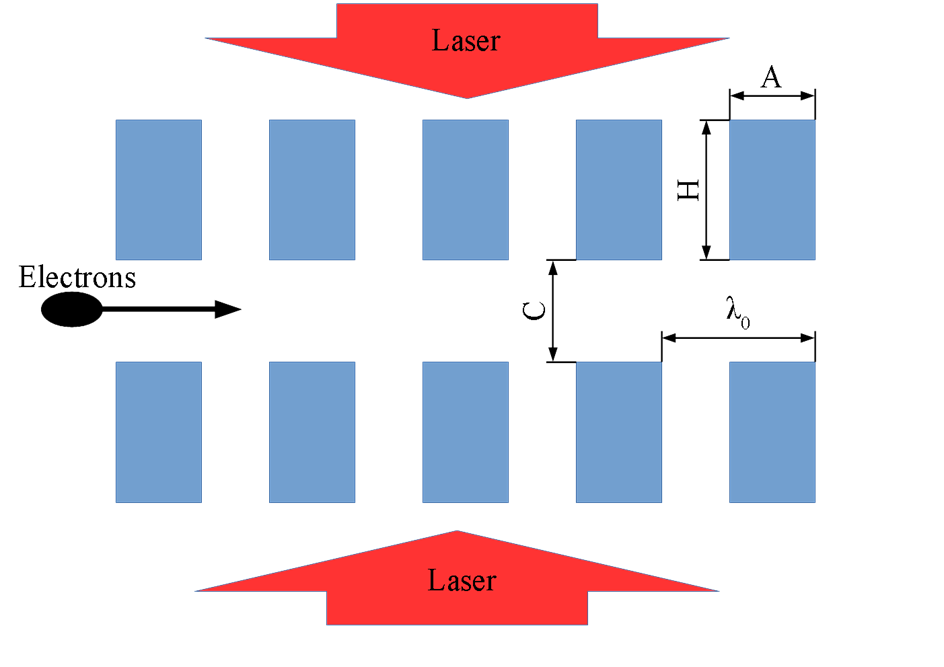}
   \caption{Schematic of the pillar grating type DLA with parameters $\lambda_{0} = \SI{2}{\micro\m}$, $A = 0.5 * \lambda_{0}$, $C = 0.37 * \lambda_{0}$, $H = 0.87 * \lambda_{0}$.}
   \label{fig:DLA_schematic}
\end{figure}

The particle-in-cell code from CST Studio Suite 2017 was used for this simulation. A grating type pillar structure with 50 periods of fused silica illuminated from both sides was simulated (see figure \ref{fig:DLA_schematic}). The total laser pulse energy is \SI{14}{\micro\J}. Spatially a Gaussian intensity profile of the laser is assumed. The electron beam is assumed to be collimated horizontally to the gap size of the DLA and to a full width of \SI{10}{\micro\m} in vertical direction allowing only small transmission of around \SI{25}{\femto\coulomb}. The beam size does not change significantly along the structure. This should still be detectable with an intensified CCD camera setup \cite{Hachmann2016}.

\begin{figure}[!htb]
   \centering
   \includegraphics*[width=78mm]{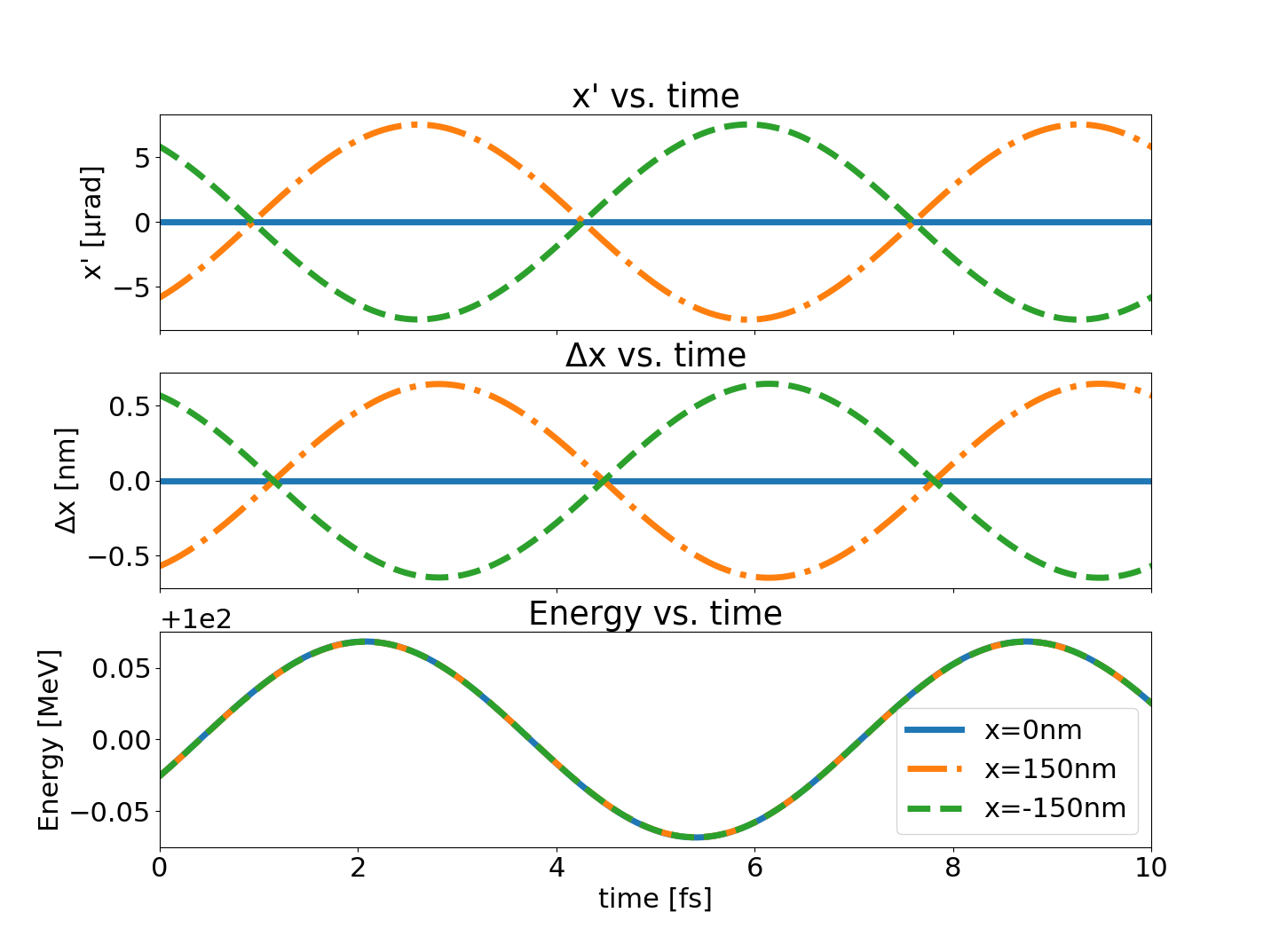}
   \caption{Shown is the data of three test particle beams from point sources at the center of the structure gap and at a horizontal offsets from the center of \SI{150}{\nano\m}. The first plot shows the horizontal transverse kick of the particles. The second plot shows the horizontal transverse offset from the starting position. And the third shows the particle energy. All show the  corresponding difference before and after the structure dependent on the injection time zoomed to the period of maximum achievable energy gain.}
   \label{fig:TestBeams}
\end{figure}

At first zero emittance test beams comprised of single electrons emitted at each timestep from point sources were simulated. Figure \ref{fig:TestBeams} shows the dependency of various particle parameters after the DLA interaction on the injection time. As expected the accelerating electron-to-laser-phase shows transverse defocusing.

At the phase of the highest achievable energy gain the bunch from the ARES simulation was injected. The same particle distribution was propagated using the in-house code.

	
\section{Results}

\begin{figure}[!htb]
   \centering
   \includegraphics*[width=78mm]{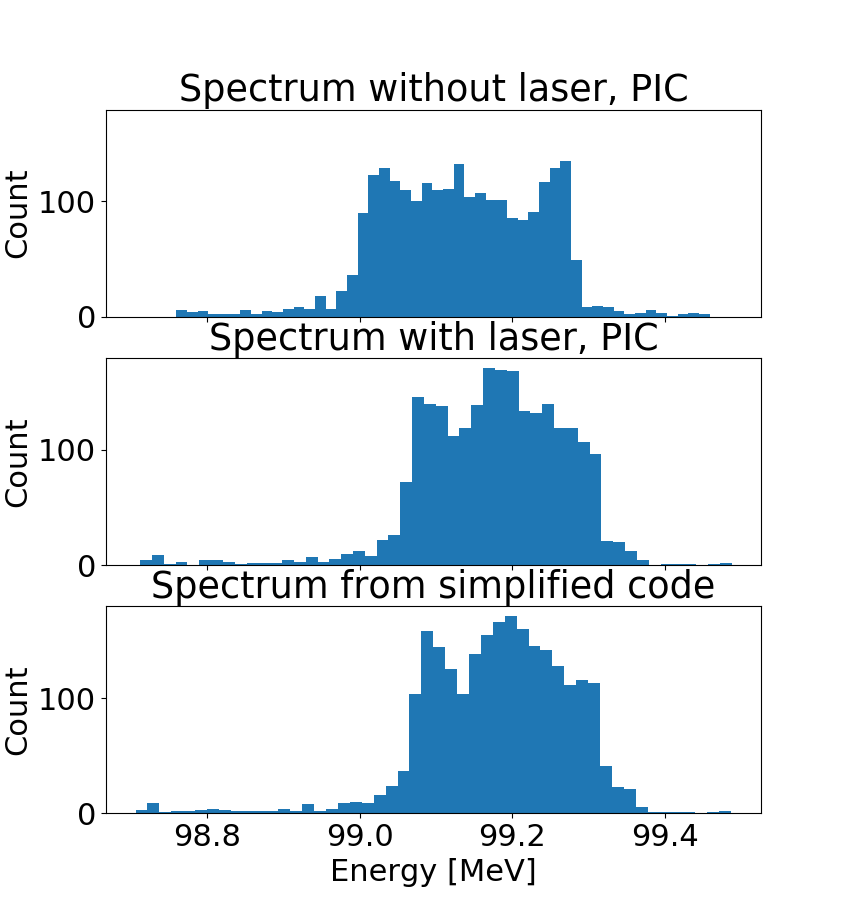}
   \caption{Shown are the histograms of the particle energy distribution after the DLA. The first histogram is without the laser field. The second one is from the PIC simulation with laser field. And the third one shows the spectrum from the in-house code.}
   \label{fig:Spectra}
\end{figure}

In figure \ref{fig:Spectra} the energy spectra of the bunch after the DLA are shown with laser illumination and without. The interaction with the structure due to wakefields and beam loading are not visible in the spectrum, which is to be expected due to the symmetry of the beam and the low charge. This is most likely due to the low bunch charge. The spectrum with the laser field shows a mean net-gain in particle energy of \SI{43.2}{keV}. This corresponds to an acceleration gradient of \SI{432}{MeV\per \m}.

The third histogram in figure \ref{fig:Spectra} shows the result from the in-house code, which is in good agreement with the full PIC simulation. The transverse phase spaces are virtually the same in all three scenarios. The short propagation distance allows for no conclusion about the accuracy of the code with respect to the PIC simulation.

It is important to mention the arrival time jitter of the ARES linac, which is expected to be in the order of \SI{10}{\femto\s}. This means that the actual bunches will be injected at a random laser to electron phase. With a maximum repetition rate of \SI{50}{Hz} sorting of the recorded spectra is feasible to look at the single shot spectra. Methods to mitigate the arrival time jitter are under investigation \cite{Mayet:IPAC2017-WEPVA006}.


\section{Outlook}

The simulations have to be extended to include possible laser system upgrades to show the full potential of the setup. Additional conventional or permanent magnet based focusing of the electron beam into the DLA should be investigated to increase electron beam transmission. Methods to mitigate the arrival time jitter have to be developed to show stable net acceleration and to scan the electron to laser phase to fully characterize DLA structures. Optimization of the DLA structures with respect to acceleration gradients and transverse beam manipulation are underway.

	
\section{Conclusion}

It was shown that under the assumed conditions net-acceleration in a DLA structure is possible at the ARES linac with only minor modifications to the beam line. Such an experiment has not been conducted so far, to the best of our knowledge. Furthermore it was shown that the in-house code can be used to model the electron-DLA-interaction under the assumed conditions.

	
\section{Acknowledgments}
	
This research was conducted under the Accelerator on a CHip International Program (ACHIP) funded by the Gordon and Betty Moore Foundation via grant GBMF4744.
\newpage
\section*{References}

\bibliographystyle{unsrtnat}
\bibliography{\jobname}

\begin{thebibliography}{12}
\providecommand{\natexlab}[1]{#1}
\providecommand{\url}[1]{\texttt{#1}}
\expandafter\ifx\csname urlstyle\endcsname\relax
  \providecommand{\doi}[1]{doi: #1}\else
  \providecommand{\doi}{doi: \begingroup \urlstyle{rm}\Url}\fi

\bibitem[et. al.(2017{\natexlab{a}})]{ACHIP}
Rasmus~Ischebeck et. al.
\newblock The accelerator-on-a-chip international program.
\newblock \emph{NIMA Proceedings}, 2017{\natexlab{a}}.

\bibitem[England et~al.(2014)England, Noble, Bane, Dowell, Ng, Spencer,
  Tantawi, Wu, Byer, Peralta, et~al.]{England2014}
R~Joel England, Robert~J Noble, Karl Bane, David~H Dowell, Cho-Kuen Ng, James~E
  Spencer, Sami Tantawi, Ziran Wu, Robert~L Byer, Edgar Peralta, et~al.
\newblock Dielectric laser accelerators.
\newblock \emph{Reviews of Modern Physics}, 86\penalty0 (4):\penalty0 1337,
  2014.

\bibitem[AXS()]{AXSIS}
Axsis.
\newblock URL \url{@Misc{CST, title = {CST}, url = {https://www.cst.com/}, }}.

\bibitem[et. al.(2017{\natexlab{b}})]{Ulrich1}
Ulrich~Dorda et. al.
\newblock Status and objectives of the dedicated accelerator r\&d facility
  "sinbad" at desy.
\newblock \emph{NIMA Proceedings}, 2017{\natexlab{b}}.

\bibitem[CST()]{CST}
Cst.
\newblock URL \url{https://www.cst.com/}.

\bibitem[Floettmann()]{ASTRA}
Klaus Floettmann.
\newblock Astra.
\newblock URL \url{http://www.desy.de/~mpyflo/}.

\bibitem[et. al.(2017{\natexlab{c}})]{Kuropka2017}
Willi~Kuropka et. al.
\newblock Simulation of a many period dielectric grating-based electron
  accelerator.
\newblock \emph{IPAC2017 - Proceedings}, 2017{\natexlab{c}}.

\bibitem[Mayet et~al.(2017{\natexlab{a}})Mayet, Aßmann, Dorda, Kuropka,
  Marchetti, and Zhu]{Mayet:IPAC2017-WEPVA007}
F.~Mayet, R.W. Aßmann, U.~Dorda, W.~Kuropka, B.~Marchetti, and J.~Zhu.
\newblock {S}imulations and {P}lans for a {D}ielectric {L}aser {A}cceleration
  {E}xperiment at {SINBAD}.
\newblock In \emph{Proc. of International Particle Accelerator Conference
  (IPAC'17), Copenhagen, Denmark, 14th-19th May, 2017}, number~8 in
  International Particle Accelerator Conference, pages 3264--3267, Geneva,
  Switzerland, May 2017{\natexlab{a}}. JACoW.
\newblock ISBN 978-3-95450-182-3.
\newblock \doi{https://doi.org/10.18429/JACoW-IPAC2017-WEPVA007}.
\newblock URL \url{http://jacow.org/ipac2017/papers/wepva007.pdf}.
\newblock https://doi.org/10.18429/JACoW-IPAC2017-WEPVA007.

\bibitem[et. al.(2017{\natexlab{d}})]{FrankEAAC1}
Frank~Mayet et. al.
\newblock Simulations and plans for possible dla experiments at sinbad.
\newblock \emph{NIMA Proceedings}, 2017{\natexlab{d}}.

\bibitem[Jia et~al.(2006)Jia, Chen, Huang, Zhao, Li, Xu, Sun, Feng, Li, Wang,
  et~al.]{Jia2006}
TQ~Jia, HX~Chen, M~Huang, FL~Zhao, XX~Li, SZ~Xu, HY~Sun, DH~Feng, CB~Li,
  XF~Wang, et~al.
\newblock Ultraviolet-infrared femtosecond laser-induced damage in fused silica
  and caf 2 crystals.
\newblock \emph{Physical Review B}, 73\penalty0 (5):\penalty0 054105, 2006.

\bibitem[{Hachmann} and {Fl{\"o}ttmann}(2016)]{Hachmann2016}
M.~{Hachmann} and K.~{Fl{\"o}ttmann}.
\newblock {Measurement of ultra low transverse emittance at REGAE}.
\newblock \emph{Nuclear Instruments and Methods in Physics Research A},
  829:\penalty0 318--320, September 2016.
\newblock \doi{10.1016/j.nima.2016.01.065}.

\bibitem[Mayet et~al.(2017{\natexlab{b}})]{Mayet:IPAC2017-WEPVA006}
F.~Mayet et~al.
\newblock {A} {C}oncept for {P}hase{-S}ynchronous {A}cceleration of
  {M}icrobunch {T}rains in {DLA} {S}tructures at {SINBAD}.
\newblock In \emph{Proc. of International Particle Accelerator Conference
  (IPAC'17), Copenhagen, Denmark, 14th-19th May, 2017}, number~8 in
  International Particle Accelerator Conference, pages 3260--3263, Geneva,
  Switzerland, May 2017{\natexlab{b}}. JACoW.
\newblock ISBN 978-3-95450-182-3.
\newblock \doi{https://doi.org/10.18429/JACoW-IPAC2017-WEPVA006}.
\newblock URL \url{http://jacow.org/ipac2017/papers/wepva006.pdf}.
\newblock https://doi.org/10.18429/JACoW-IPAC2017-WEPVA006.

\end{thebibliography}

\end{document}